# Surface Finishing and Coatings for Accelerator Vacuum Applications

*M. Taborelli*
CERN, Geneva, Switzerland

**Abstract**
The main applications and techniques of thin film coating and plating on vacuum components for particle accelerators are reviewed through significative examples.

## 1  Introduction

Surface finishing and coatings enable to modify and tune the properties of the surface of a material for a specific application. In many cases to fulfil all requirements set by an application it is easier to separate the functions by using for instance a structural bulk material and modifying only its surface to provide the other functions. For the specific case of vacuum and Ultra High Vacuum (UHV) systems in accelerators [1], the surface of a material influences the thermal outgassing and particle induced outgassing (electron-, photon-, ion-stimulated desorption), the electron multiplication (electron cloud and multipacting), the performance of superconducting thin films in the RF domain, the impedance of a beam-pipe, the resistance to sparking and the corrosion resistance of the device. In manufacturing technology the surface controls, among others, the wettability of brazing alloys and the adhesion of thin films to the substrate. In practical situations, the material is not ending with a defect-free geometrical plane and the concept of surface is no longer restricted to the topmost atomic layer, but concerns rather a top layer of few nano-meters or even microns thickness.

The surface, for a simple geometrical reason, is the most exposed part of a component and after the surface finishing it shall be protected. More important is to realise that a "finishing" shall be the last treatment to be applied to a product in the manufacturing sequence and no further handling or manufacturing step should be added.

These review presents some typical methods of treatment of the surface of particle accelerator components, namely cleaning, electroplating and coating.

## 2  Cleaning

Cleaning is the most basic surface finishing of a UHV part, but its importance and difficulty should not be underestimated. The fabrication processes, as mechanical machining, leave residues of lubricants, fingerprints, marker traces and chips on the surface. This contamination leaves the parts in an undefined state and induces strong outgassing of hydrocarbon-based compounds when inserted in vacuum. Any component, which is produced for insertion in UHV (typical pressure below $10^{-7}$ mbar or $10^{-5}$ Pa) must undergo a precision cleaning process to remove those residues and afterwards it must be manipulated, packaged and stored with the necessary precaution. The cleaning occurs in most of the cases in a water based detergent bath or in an organic solvent. More rarely, a dry cleaning based on air or oxygen plasma is applied. A summary of cleaning techniques applied to components designed for use in UHV, storage effects and the related methods to characterise the surface cleanliness is presented in the report of a previous CERN Accelerator School [ 2, 3] and is not repeated here.

It is important to underline that the possibility to easily clean the parts for UHV application must be included from the beginning in the design. Traps and pockets (non-penetrated welds, voids, etc…), which cannot be rinsed or easily reached by a cleaning fluid should be avoided and small diameter blind tapped holes should be limited. Grinding, brushing and sandblasting resulting in rough surfaces must be restricted and, if unavoidable, the materials must be carefully selected to avoid embedding inclusions of the abrasive or leave porosities. Stickers and marker traces on functional surfaces shall be avoided from the beginning. Finally, excessive amounts of cutting and lubricating fluids should be removed as soon as possible from the parts even before precision cleaning and solid lubricants (graphite, $MoS_2$, BN, etc…) shall be avoided to simplify the subsequent precision cleaning, since their dust provokes bath pollution.

The cleaned surface has in general a high surface energy and during air exposure can be partly re-contaminated by airborne adsorbates. For some metals, like copper, the cleaned surface becomes reactive and air exposure can induce further oxidation, formation of hydroxide in presence of humidity and discoloration. The necessity of a further treatment helping to stabilize the surface (often called passivation) shall be evaluated depending on the application and required properties.

## 3    Electroplating

Electroplating or electrochemical plating or galvanic deposition is a process frequently applied in industry to modify the surface of a part either to protect it from corrosion or to tune other surface properties [4, 5]. For the case of particle accelerators, the most common applications are the following:
- improving electrical surface conductivity on vacuum pipes and other surfaces facing the beam (reduce impedance, enable flow of the image current) (Cu),
- enabling mobile electrical contacts in UHV and avoiding sticking (Au, Ag, Rh/Cu; in general, hard on soft hinders sticking and gives a large contact surface),
- preventing cold welding, seizing, (Ag on nuts or bolts of flanges),
- providing a wetting layer for brazing filler metals (Ni on StSt) or provide a diffusion barrier (Cu on Glidcop) to avoid dilution of the filler in the bulk,
- protecting against oxidation or corrosion, preserving reflectivity or emissivity.

When the same method is used to form a standalone object, it gets the name of electroforming.

Very schematically (Fig. 1) the electroplating process requires to immerse the part in an electrolyte bath with a solution containing ions of the metal to be deposited. Those ions can be provided either by an anode made of the same material to be deposited (soluble anode) or are dissolved from a salt in the bath and the anode has only the function of establishing the electrical potential (insoluble anode). The cathode is the part to be plated and is connected to the negative pole of a current source to attract the positive metal ions from the bath. At the cathode the metal ions are reduced to metal (valence zero) and coat the part. In case of soluble anode, the element of the anode is oxidised, and ions are dissolved. For an insoluble anode the current balance is guaranteed by the production of protons from water. The transferred charge (the integral of current vs time) is proportional to the amount of material which is deposited (Faraday law), weighted with the reaction yield. The ions can be stabilised in the solution by a specific pH and other specific additives. The smoothness of the deposited layer can be ensured by additives as well, but for vacuum applications the addition of organic additives might enhance outgassing and the single cases must be validated on preliminary test-pieces.

The most frequent substrates to be plated for accelerator applications are copper, stainless steel and more seldom aluminium. Metals which build a very stable surface oxide upon air exposure ($Cr_2O_3$, $TiO_2$, $ZrO_2$, $Al_2O_3$….) cannot be used directly as substrates, since the stable oxide layer prevents adhesion of the plated layer. In the cases of stainless steel (with $Cr_2O_3$ on the surface) and aluminium alloys ($Al_2O_3$) some specific pre-treatments can prepare the surface for electroplating. For stainless steel it is common to use a pre-layer of Ni or Au (used especially when magnetic properties are an issue) and for aluminium a chemical double zincate treatment is applied. [6]. The substrate must be a conductor, since the amount of deposited material is proportional to the electric charge passed through the circuit. However, electroplating can also be performed on some insulators, as polymers, glass or ceramics, after a suitable surface chemical pretreatment [7] or coating with a thin metallic layer (see next section) to establish a sufficient electrical conductivity to start the electrochemical process (after a certain thickness the deposited layer can carry itself the electric current). Such solutions are not straightforward and need preliminary testing.

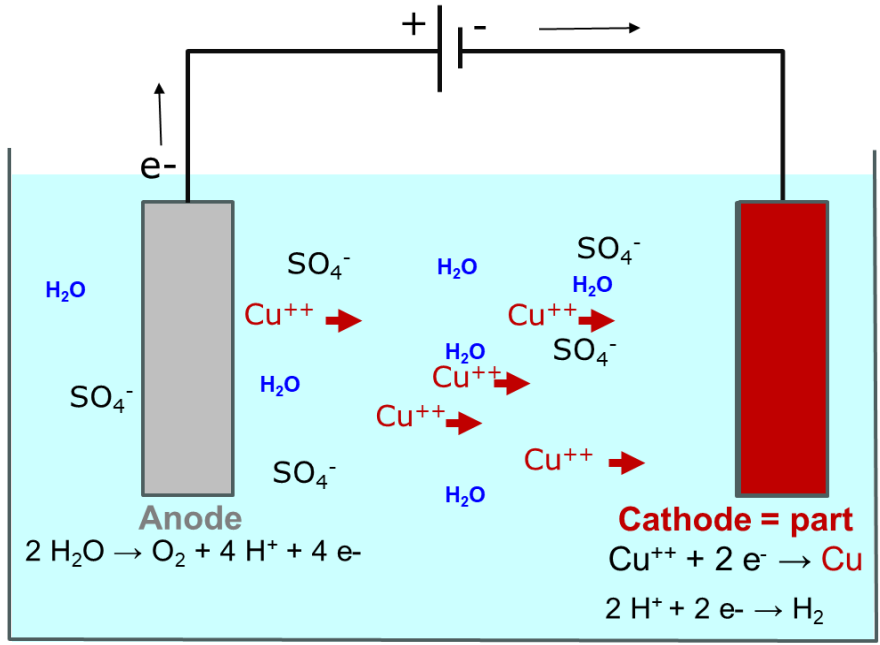

**Fig. 1:** Electroplating with insoluble anode

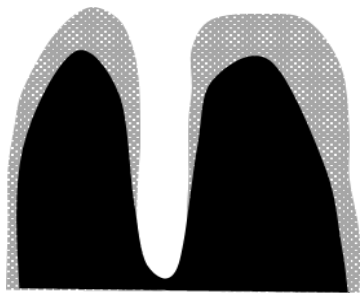

**Fig. 2:** Schematics of the difference in plating thickness on protrusions and valleys (plated layer in grey)

Plating complex shapes introduces a further difficulty, since it results in a non-uniform thickness for the deposited layer. The ions in the bath are guided by the spatial distribution of the electric field and resulting current lines so that protrusions and valleys will not be covered with the same amount of material. The most evident example are the convolutions of a bellow (Fig. 2) where the protruding wave will receive more deposited material than the deep valleys of the convolutions; similarly, the edges of a flat plate will receive a larger plating thickness than the central flat part (edge effect). A preliminary test-piece with a representative geometry or a simulation [8] of the process in commercial multi-physics software packages like COMSOL® can reveal the extent of such effects and will assess whether the process will be able to deliver a part fulfilling the specification. In any case, as for the cleaning process, the part shall be designed from the beginning by taking into consideration the final plating process. The materials constituting the part shall be compatible with the plating bath chemicals if the part must be completely immersed. The shape shall enable the effective removal by rinsing of specific corrosive chemicals (as HCl which is corrosive for stainless steel and is used for Ni plating), which will be in contact with the full piece. Finally, it sounds very basic, but it is worth to note that the availability of the suitable infrastructure (plating bath size for instance) in industry or research institutes shall be evaluated already at the design phase.

Electroforming is a manufacturing process to build an entire component and enables to deposit thicker layers by following the same principle as electroplating. The substrate which is covered by the deposited layer is a sacrificial mandrel [9,10]. The latter option can be used when it is easier to machine a mandrel with the shape of the future empty space (like the empty part of tube or a cooling circuit or a cavity) than to machine the real object. The mandrel is finally dissolved by chemical means or exploiting different thermal expansion or melting. In the case of accelerator applications this technique has shown the possibility to produce copper thin walled chambers of high aspect ratio (few mm diameter and one or two meters length, Fig. 3) [9] and RF cavities of moderate size (1.3 GHz for instance) [10]. The mandrel can also be coated first by using other deposition techniques and in a second phase the electrodeposited thick layer is built on top. In this way, for instance, a very small diameter vacuum pipe can be coated on the inner surface with a suitable functional material [9].

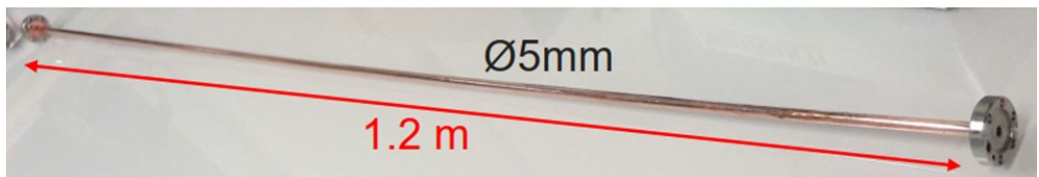

**Fig. 3:** Electrodeposited vacuum chamber with integrated CF DN16 flanges.

## 4      Thin films coatings

In this section we consider films deposited by Physical Vapour Deposition (PVD) in a range of thickness below 2-3 micrometers. Depending on the material, larger thickness might be achievable but shall be tested with respect to adhesion of the coating, since the stress of the film is related to its thickness. There are many applications in accelerator technology exploiting surface modification by thin film coating and the most prominent are the following:
- reducing the material outgassing (thermal and particle induced) and providing pumping action by non-evaporable getters,
- reducing the Secondary Electron Yield (SEY) to mitigate electron cloud or multipacting,
- improving electrical surface conductivity on vacuum pipes, other surfaces facing the beam (reduce impedance, enable flow of the image current) (Cu) or electrical contacts,
- providing a thin superconducting surface layer in RF accelerating cavities or beampipes (impedance),
- providing thin layers of specific materials as targets or wavelength shifters.

The last two cases will not be discussed in the following and some examples are in Refs. [11-14].
A detailed presentation on the coating techniques applied to particle accelerators can be found in Ref. [15]. Magnetron sputtering or glow discharge magnetron sputtering is the method of PVD, which is most commonly used. Briefly, the parts to be coated are inserted in a dedicated vacuum chamber, or in case of a beam pipe the chamber is the part itself (Fig. 4). The chamber is pumped down to UHV level, if necessary, by applying a bake-out. For most of the coatings discussed below, it is mandatory to reach UHV level to guarantee the performance of the films and their adhesion. This clearly implies that all parts involved are correctly cleaned (see previous sections) and are UHV compatible. An inert gas is injected in the chamber in a pressure range from 1e-3 to 1e-1 mbar and the High-Power Impulse Magnetron Sputtering (HiPIMS), is entering the stage and providing the advantage of more conformal coating and an increase in the energy of the deposited atoms. In this technique the high current provided during a short pulse increases the fraction of ionized species emitted from the target, which can hence be accelerated and directed to the substrate by an electric filed generated through an applied negative bias voltage with respect to ground. [16]

In case of vacuum chambers to be coated on the inner surface, the material substrates are generally stainless steel, copper and its alloys (OFE Copper C10100, OFS copper C10700), aluminium alloys (series 6000 and 5000, more rarely 2000) or pure beryllium in the very special case of interaction point vacuum chambers. For other components a wide range of materials include also graphite, glass, ceramics or polymers. Depending on the applications the coating can also be deposited on top or below an electroplated layer. For each case an adequate substrate preparation process, often beyond cleaning, shall be foreseen (etching, chemical passivation etc...) to guarantee the adhesion of the deposited coating. Another processing step to enhance adhesion is the removal of the surface oxide by a so/called inverse ion sputtering or ion etching, either with the help of a temporary anode or with a suitable ion source. For substrate materials which are prone to react in air (typically copper and its alloys) and build fragile oxides or hydroxides it is necessary to limit the air exposure after surface preparation. According to this rationale the planning should minimize the time between the surface preparation and the thin film coating to reduce the air exposure time between the two processes to one or two days. In this sense having the two facilities one close to the other simplifies the task.

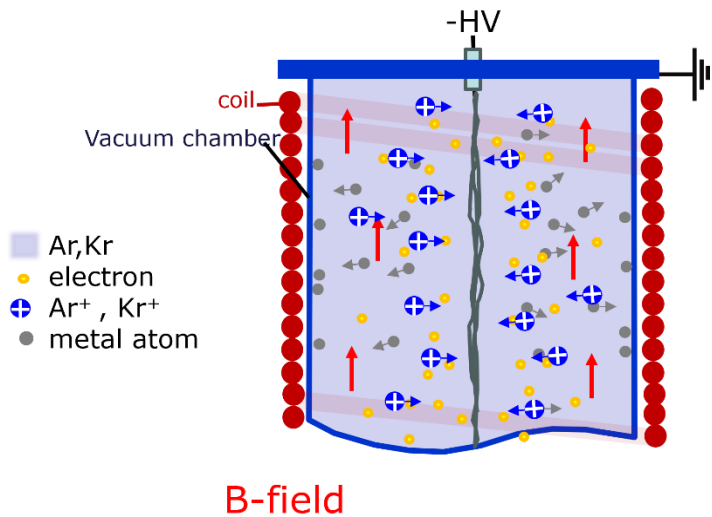

**Fig. 4:** Scheme of DC magnetron sputtering system for the coating of the inner wall of a tube shaped vacuum chamber.

In addition to the material itself, some steps in the manufacturing may weaken locally the adhesion of the coating and this shall be considered in the design or in the preparation before coating: in most cases an adapted chemical etching which "erases the memory" of the manufactured surface is necessary and sufficient. Few examples from the experience of the author are listed in the following. Electropolishing of stainless steel, which is often applied to reduce the effective outgassing surface, has been shown to hinder the adhesion and a mild etching can solve the problem. Copper chambers are produced by extrusion when an excellent dimensional accuracy of the diameter is required; however, when the extrusion mandrel is worn-out it induces the risk of embedding loose copper chips on the surface, which will detach spontaneously (for instance during bake-out) together with the deposited coating. In this case the etching of several tenths of microns of the surface are required. Machining with a lathe of the surface of stainless-steel flanges has provoked in few cases difficulty in the adhesion of the subsequent coating, possibly because of overhanging burrs along the tool path. The issue with lathe machining was also encountered on the inner surface of thin-walled aluminium alloy pipes, without any clear visible feature which could easily explain the effect. Welding lines are by definition regions where the material is different from the original bulk and often coating adhesion. Finally, a very specific issues has been encountered on 316LN stainless steel after vacuum firing at 950C, with the surface segregation of boron nitride chips, which are loose particles not adhering to the surface and exhibiting a low surface energy [2].

### 4.1 Non-Evaporable Getters

Thin films of non-evaporable getter (NEG) materials are commonly used in accelerator vacuum systems worldwide. Several reviews and presentations are available in literature [17]. The first large scale deployment (6 km length) in the long straight sections of LHC [18] demonstrated the robustness and effectiveness of this technology which has been later adopted in several synchrotron machines up to the last generation [19-22].

In accelerator technology the function of the getter film on the surface of a vacuum chamber is to provide a distributed pumping on the entire surface after thermal activation. The pumping occurs because of the high reactivity of the NEG activated surface, which can bind molecules like $H_2$, CO, $CO_2$ – all typical in the residual pressure of a baked system - as well as $H_2O$ and $O_2$. Hydrocarbons, like $CH_4$, and noble gases are not pumped by the NEG coating. The fact that the getter has a high solubility for hydrogen enables it to reduce the outgassing of this gas from the underlying substrate. The electron- [23], photon- [24] and ion-stimulated [25] outgassing are reduced as well by orders of magnitudes compared to an uncoated surface of stainless steel, aluminium or copper. Last, but not least the activated TiZrV NEG surface has an SEY maximum slightly above 1 (we call in the following SEY maximum the maximum of the SEY as a function of the energy of impinging electrons) [26] and can be used to mitigate electron cloud. All the properties mentioned above are obtained only after a process of thermal activation which requires to evacuate the chamber to a high vacuum level before rising the temperature to about 200C for 24 hours and let it cool down again. The activation enables to diffuse oxygen of the oxide and adsorbates from the surface to the bulk of the coating and provides the clean surface which is reactive and can bind several types of molecules by chemisorption. The activation can be repeated after each air venting and the functional lifetime of the coating is given by the combination of oxygen solubility, number of venting/activation cycles and film thickness.

At present the most used alloy is the ternary TiZrV deposited by DC magnetron sputtering with a cathode built from intertwisted wires of the 3 elements (long chambers) or from a bulk alloy plate (planar magnetron). The long chambers are completely inserted in a solenoid coil or a shorter mobile solenoid is swept over the chamber length to optimize the film thickness distribution. Beyond TiZrV, similar alloys have been reported in literature [27] but are not deployed at large scale, yet. For vacuum chambers of complex geometry, as the combined electron and photon chambers with relatively small diameter in the 4[th] generation synchrotrons, the approach for the coating can include multiple cathodes and ad hoc designed spacers [20, 22].

## 4.2 Films to lower the secondary electron yield (SEY) without activation

As mentioned above, the NEG coating produces a surface with a maximum SEY slightly above 1 after thermal activation and enables to mitigate electron cloud when applied to the inner surface of beam vacuum chambers. An alternative solution must be provided for all cases where it is not possible to perform a thermal activation on a vacuum device, for instance because the vacuum chamber is clamped in the massive iron yoke of a magnet or because the device is in a cryogenic section of the accelerator. In RF couplers or RF windows, an SEY well below that of the insulating material (often alumina, which has an SEY maximum around 7) integrated in the device is necessary to avoid multipacting: clearly in this case the task is easier, since an SEY close to 2 is already beneficial, whereas to mitigate electron cloud in a beampipe the required value is close to 1.

Titanium nitride, TiN, is often quoted for application in vacuum chambers [28] and in RF devices [29, 30]. Stoichiometric TiN is difficult to produce, since it requires a precise and homogeneous flow of $N_2$ in the reactive magnetron sputtering system, which is not easily achievable for instance on long beampipes. Even if the material has an SEY close to 1 after deposition, the value for smooth films rises to 1.4 to 2.5 upon air exposure, depending on the deposition parameters. This is due to the binding energy with Ti which is higher for O than for N; as a result a partial substitution of surface atoms and formation of a $TiN_xO_y$ layer occurs [31] leading to an SEY in the range mentioned above. This is still not an issue for a TiN coated alumina in RF application, for instance, since the SEY is anyway strongly decreased compared with the bare insulator. However, it is not acceptable for a coating aimed at mitigating electron cloud, where the target SEY maximum must be close to 1. On a TiN air exposed surface, such a low SEY maximum close to 1 can only be reached after conditioning (called also scrubbing), which consist in irradiating the surface with electrons of moderate energy (200 eV-1keV) and can be obtained through operation at reduced beam intensity, but still sufficient to provoke electron cloud) [32]. The process implies obviously some cost due to the machine operation with a beam which is useless for physics.

For RF application a thin layer of pure Ti provides a valid alternative and is much easier to apply. The thin layer which is aimed at suppressing multipacting should not spoil the function of the underlying insulator and therefor the Ti film shall remain very thin. An established strategy [33] is to tune the thickness by monitoring the resistivity under vacuum during the magnetron sputtering deposition and keeping in mind its increase upon air exposure. In this way a resistivity square in the range 5- $10^6$ Ohm-square can be obtained.

Carbon has intrinsically a low SEY and for the mitigation of electron-cloud in beampipes, a technical solution based on amorphous carbon thin films has been developed [34, 35] initially for the Super Proton Synchrotron (SPS) magnet chambers at CERN. These films provide a maximum SEY of 1-1.1 without the need of bake-out even after prolonged air exposure, when protected just from gross contamination and dust. They were successfully deployed to upgrade more than 100 quadrupole magnets and short straight sections of the SPS [36], where part of the vacuum chambers is trapped in the iron yoke. The necessary thickness for the target maximum SEY is about 50 nm on smooth stainless steel and a safety margin of a factor 2 in thickness has shown to be reliable to keep the SEY very close to 1 in a large production. The coating has been also applied to short room temperature (replacing solenoids to suppress electron cloud) and cryogenic sections of LHC [37]. In the latter case the mitigation of electron cloud has been demonstrated by the drastic decease of the beam induced heat load [38]. Amorphous carbon thin films are also the baseline for a large number of new devices, from the beampipes (beamscreens) of the upgraded straight sections to be installed for the HL-LHC project, the upgrade of RHIC in the context of the EIC project [39] and last but not least, the upgrade of the LHC arcs to reduce the electron cloud induced heat load.

Amorphous carbon has been successfully deployed on stainless steel, copper, copper-beryllium and aluminium and there are no obvious limits for applications on other metallic materials. A peculiar case is the coating on quartz for a future LHCb experiment. [40]. The surface of new components can be prepared by conventional wet chemistry means. Instead, for refurbishment of beam pipes which are already installed and cannot be cleaned by wet chemistry (especially for copper) a preliminary thin (50 nm) of titanium coating is necessary to ensure the carbon film adhesion.

The coating methods with amorphous carbon in accelerators were selected and developed in tight connection with each application with its specific constraints. DC magnetron is the preferred option in terms of effectiveness (process time) and is applied for open parts with planar graphite targets. For tube geometry a dedicated mobile sputtering device with permanent magnets and cylindrical targets has shown excellent results [37]. However, the very first large scale deployment in the SPS was based on hollow cathode coating device. The vacuum chambers are trapped in the magnet yoke and the magnetic field (dipoles, quadrupoles and multipoles) is never on the chamber axis, as it would be suitable for DC magnetron with a carbon rod target along the axis. Therefore, the option of coating in hollow-cathode configuration without magnetic field was chosen [36]. It is worth to note that for each new geometry a development phase shall be considered, with tests in different options. About 10 years elapsed from the first coated plate of some square centimetres exhibiting a maximum SEY of 1 with satisfactory adhesion to the large deployment in SPS.

### 4.3  Films to increase surface conductivity

The electrical resistivity of the surfaces facing the beam contribute to the global balance of the machine impedance. This contribution is in general more critical when the surface is close to the beam path, like for collimators. This aspect is not an issue when the jaws are made of metallic materials, but gets relevant at the machine positions where low density materials, typically graphite and carbon based composites [41], are necessary. In such cases a thin metallic coating can improve the machine performance.

From the functional point of view the thickness of the coating shall be defined on one hand considering the skin depth for the suited electromagnetic frequency range and on the other hand keeping in mind that coating adhesion is more difficult when thickness increases. The electrical conductivity of a thin film of an element or compound is generally lower than the one of the corresponding bulk due to the intrinsic smaller grain size and higher density of defects. This shall be considered as well when defining the target thickness for the application.

The typical material for such coatings is copper [42] with its high electrical conductivity, however, for beam intersecting devices where beam impact and resistance to high temperatures was required, molybdenum was considered to be the best choice [43, 44].

A serious balance between risks and advantages of a coating with respect to the bare substrate shall be made case by case on such substrates. The main difficulty for the carbon-based substrates, as graphite, is the fragility of the layered material and its mechanical resistance which is often weaker than the adhesion of the coating. The substrate surface presents often loosely bound fragments, which are prone to detach together with the coating during operation. Extensive tests shall be performed comparing cleaning and surface preparation methods (solvent- or water-based, with or without ultrasonic agitation in the bath, $CO_2$ jet, etc...) to obtain a reasonably robust final product.

## 5    Quality control

The quality control depends on the type of surface finishing (cleaning, etching, electro-plating, coating) and on the desired functional properties. A first gross visual examination can reveal defects and non-conformities for most of the treatments, but a visible non-uniformity does not always imply a reduced performance of the component.

For cleaning, surface chemical analysis as for instance X-ray Photoemission Spectroscopy, is the most powerful tool. In most of the cases this cannot be applied directly on the components, which do not fit in the analysis chamber, and is performed on a witness sample following the same treatment. A more detailed description of cleanliness assessment methods after cleaning is provided in Ref. [2].

For electroplating the typical properties which can be tested are the thickness and the adhesion. The thickness can be measured in a non-destructive way by X-Ray Fluorescence (XRF) when it does not exceed about 20 um (this value is for copper and it depends on the atomic number of the layer material). Portable XRF devices enable local measurements for parts which offer sufficient access and in other cases the control must be performed on witness samples. For thicker films, eddy current probes are better adapted. Finally for very thick films in the range of 100 um the usual mechanical metrology tools can be used. An average thickness can also be obtained by considering the uptake of weight. Adhesion testing is generally destructive. A very practical method (ASTM D3359) is based on a multi-blade cross hatch adhesion cutter, which can engrave a checkerboard pattern on the plated layer. A calibrated adhesive tape is then glued on the checkerboard engraved area and ripped off; the resulting pattern of damages to the layer enables to classify the adhesion degree. Other properties, as the electrical conductivity can be assessed by specific methods [44, 45].

For PVD coatings the thickness and adhesion are typical properties to be tested as well. The XRF method is valid in the typical range of thickness of these layers and can be applied when the surface is easily accessible. For films below 100 nm an observation in Scanning Electron cross section in Focussed Ion Beam For a test of the thickness distribution in complex shapes there is no other alternative than a mock-up which can be sacrificed or can host a series of samples. Adhesion can be tested by the cross-hatch method mentioned above. A more quantitative alternative is a pull test (for instance ASTM D4541) where a dolly is glued on the surface and is pulled off mechanically with a device which can measure the pulling force. The result is a fractured face at a measured stress (force/surface). It is important to verify whether the fracture is adhesive (between layer and substrate) or cohesive (within the substrate). In the latter case the adhesion of the coating is considered satisfactory independent on the pull-force value, since the adhesion is stronger than the strength of the substrate itself. Verification of the functional properties as pumping speed, SEY, conductivity, is only rarely possible on the real par and witness samples must be used.

# 6 Conclusions

Specific properties which are different from those of the bulk material, can obtained by adapted surface treatments. In all the cases presented above the surface treatment shall be included from the beginning in the fabrication process to guarantee the compatibility of the materials and geometries with the treatment itself. Moreover, the surface being the most exposed region of the part and being therefore prone to damages, the surface treatment should be the last process on the object before packaging and installation.